\documentstyle[sprocl,epsf]{article}

\input{psfig.sty}

\def\Journal#1#2#3#4{{#1} {\bf #2}, #3 (#4)}

\def\aas{\em A\&AS}
\def\apj{\em ApJ}
\def\apjs{\em ApJS}
\def\aap{\em A\&A}
\def\mnras{\em MNRAS}
\def\aj{\em AJ}
\def\nature{\em Nature}
\def\araa{\em ARAA}
\def\science{\em Science}
\def\pasp{\em PASP}

\def\be{\begin{equation}}
\def\ee{\end{equation}}
\def\bea{\begin{eqnarray}}
\def\eea{\end{eqnarray}}

\def\mh{{M_{\bullet}}} \def\mb{{M_{\rm bulge}}} \def\msun{{M_{\odot}}}
 \def\ms{{\mh-\sigma}} \def\vs{{v_c-\sigma}}
  \def\etal{{\it et al.\ }}
 \def\ml{{\mh-M_B}}

\def\Sec{\hbox{${}^{\prime\prime}$\llap{.}}}

 \def\kms{km s$^{-1}$}
 
\def\lae{\mathrel{<\kern-1.0em\lower0.9ex\hbox{$\sim$}}}
\def\gae{\mathrel{>\kern-1.0em\lower0.9ex\hbox{$\sim$}}}

\def\fun#1#2{\lower3.6pt\vbox{\baselineskip0pt\lineskip.9pt
  \ialign{$\mathsurround=0pt#1\hfil##\hfil$\crcr#2\crcr\sim\crcr}}}

\begin{document}

\title{BLACK HOLE DEMOGRAPHICS}

\author{LAURA FERRARESE}

\address{Rutgers, the State University of New Jersey, 136
Frelinghuysen Road, Piscataway,\\ NJ 08854, USA\\E-mail:
lff@physics.rutgers.edu}

\maketitle\abstracts{}

\section{Introduction}

It is frequently the case that revolutionary scientific ideas are
first proposed and then remain dormant for years, or sometimes
decades, before their importance is truly appreciated. Often, the
reawakening of interest is driven by new technological developments.
Such was, for instance, the case of supermassive black holes (SBHs) in
galactic nuclei.  By the mid 1960s, just a few years after the
discovery of QSOs, it was generally recognized that their energy
source must be gravitational in nature. Yet for the following three
decades the existence of SBHs  was destined to be surrounded by
skepticism. By the mid 1990s, a few years after the launch of the
Hubble Space Telescope, it was widely accepted.  Today, it is
generally agreed upon that SBHs  play a fundamental role in the
formation and evolution of their host galaxies. Freed from the burden
of having to demonstrate the very existence of supermassive black
holes, we can now begin asking more fundamental questions: how are
black holes related to their host galaxies,  how did they form, how do
they accrete, how do they evolve, and what role  do they play in the
formation of cosmic structure?

The purpose of this contribution is to review the current status of
black hole demographics. I will not address the various techniques
that are used to measure black hole masses: excellent discussions can
be found in the recent literature (e.g. Kormendy \& Richstone 1995; Ho
1999). Neither will I discuss the somewhat tumultuous events that lead
to a critical reassessment of the ``Magorrian relation''  (Magorrian
\etal 1998) and its analog for local AGNs (Wandel 1999) since  a full
description of such events can be found elsewhere (Merritt \&
Ferrarese 2001c). Instead, I will revisit the issue of black hole
demographics in  light of recent advances in the study of high
redshift QSOs (section 2), local AGNs (section 3) and local quiescent
galaxies (section 4). I will then outline the prospects for future
progress (section 5), and discuss what I believe will be the
challenges for the years to come.

\section{Black Hole Demographics: High Redshift QSOs}

The existence of SBHs in the nuclei of nearby galaxies has gained
popular consensus only in recent years. That supermassive black holes
must power QSO activity has,  however, been widely suspected since the
mid 1960s (e.g. Robinson \etal 1965).  It is therefore not surprising
that the first studies of black hole demographics were conducted, over
two decades ago, using optical counts  of high redshift QSOs. In a
seminal paper entitled ``Masses of Quasars", Andrzej Soltan (1982)
proposed a simple argument: QSO optical number counts yield a QSO
luminosity function which can be integrated to give a mean comoving
energy density in QSO light. After applying the appropriate bolometric
corrections and assuming a reasonable conversion factor of mass into
energy, Soltan concluded that the SBHs powering high redshift ($z >
0.3$)  QSOs comprise a total mass density of $\sim 5 \times 10^4$
M$_{\odot}$ Mpc$^{-3}$, each SBH having a mass of $10^8 - 10^9$
M$_{\odot}$. Soltan's arguments, which  have been employed many times
in the following years (Chokshi \& Turner 1992; Small \& Blandford
1992; Salucci \etal 1998), lead to the inescapable conclusion that
most, if not all, nearby galaxies must host dormant black holes in
their nuclei. This finding has been the main driver for SBH searches
in nearby quiescent galaxies and has kindled the interest in the
accretion crisis in nearby galactic nuclei (Fabian \& Canizares 1988),
ultimately leading to the revival of  accretion mechanisms with low
radiative efficiencies (Rees \etal 1982, Narayan \& Yi 1995).

\begin{figure}[t]
\psfig{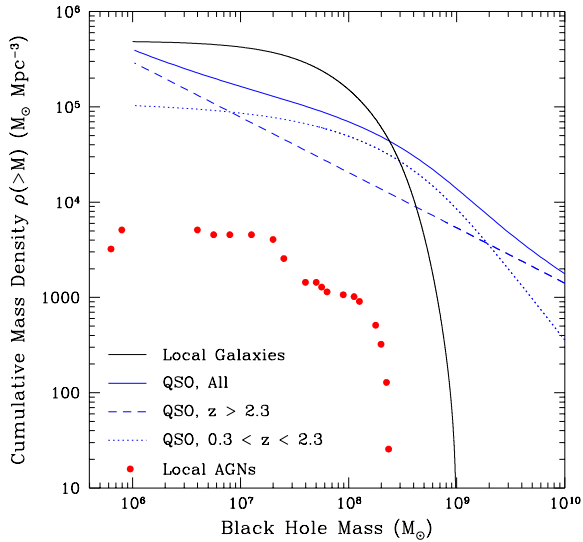}
\caption{Comparison between the black hole mass function in high
redshift QSOs (blue lines - dotted line: 2dF sample; dashed line: SDSS
sample; solid line: both samples combined); local AGNs (red) and local
quiescent galaxies (black, corresponding to the dotted black line in
Fig. 4).}
\end{figure}

Armed with recent measurements of the QSO luminosity function from the
2dF QSO Survey ($0.3 < z < 2.3$, Boyle \etal 2000) and the Sloan
Digital Sky  Survey ($3.0 < z < 5.0$, Fan \etal 2001), we are in a
position to update Soltan's results. If $\Phi(L,z)$ is the QSO
luminosity function, the cumulative mass density in SBHs which  power
QSO activity can be expressed as:

\begin{equation}
\rho_{QSO}(> M)={{K_{bol}} \over {\epsilon c^2}}
\int_0^{\infty}\int_L^{\infty}{{{L' \Phi(L',z)} \over
{H_0(1+z)\sqrt{\Omega_m(1+z)^3 + \Omega_{\Lambda}}}}}{\rm d}L' {\rm d}z
\end{equation}
where the mass accretion rate is simply
$=K_{bol}L\epsilon^{-1}c^{-2}$, with $K_{bol}$ the bolometric
correction  (from Elvis \etal 1986), and $\epsilon$ the energy
conversion coefficient (assumed equal to 0.1). An
$\Omega_\Lambda=0.0$, $\Omega_m=1.0$, $H_0 =75$ \kms~ Mpc$^{-1}$
cosmology is assumed for consistency in comparing the results with
those derived in the following sections. The cumulative mass  density
due to QSO accretion is shown in Fig. 1. It should be noted that the
magnitude limits of the 2dF and Sloan QSO surveys correspond to
Eddington limits on the SBHs masses of $4.5\times 10^7$ $\msun$ and
$7.3 \times 10^8$ $\msun$ respectively. Cumulative mass densities down
to $10^6$ $\msun$ are calculated on the (unverified) assumption that
the QSO luminosity function holds at the corresponding magnitude ($B
\sim -19$). Furthermore, the lower redshift limit of integration for
the SDSS luminosity function was pushed down to the high redshift
boundary of the 2dF survey ($z = 2.3$), although there are no QSO
luminosity functions covering the $2.3 < z < 3.0$ range. For masses
larger than $10^8$ $\msun$,  the extrapolation from $z = 3$ to $z =
2.3$ of the spatial density (e.g. Fig. 3 of Fan \etal 2001) or mass
density (Fig. 7 in these proceedings) as a function of redshift from
the SDSS joins rather smoothly the curve derived from the 2dF survey,
therefore our assumption is likely justified. However, for smaller
masses or luminosities, the SDSS mass density, extrapolated to $z \sim
2.3$, overpredicts the QSO mass density (per unit redshift) derived
from the 2dF data by an order of magnitude. Thus, it is likely that
the linear rise of the SBH cumulative mass density for the high
redshift QSOs between $10^8$ and $10^6$ $\msun$ represents an {\it
upper bound} to the real curve, which could have been overestimated by
a factor of a few (i.e., up to $\sim$ three).

In short, the cumulative mass density from the optical QSO counts due
to accretion onto high redshift QSOs ($0.3 < z < 5.0$) appears to be
in the range $(2 - 4) \times 10^5$ $\msun$ Mpc$^{-3}$.  Notice that
this estimate does not account for the possibility that sizable black
holes might have already been in place {\it before} the optically
bright phase of QSOs. Furthermore,  I have neglected the contribution
to the SBH mass density from the so called ``obscured'' or ``Type II''
QSOs, the existence of which is required to explain the observed
properties of the X-ray background. In analogy with local Seyfert 1
and Seyfert 2 galaxies, in Type II QSOs molecular material, with
column density in the neighborhood of $10^{23}$ cm$^{-2}$, completely
hides the nucleus from view at optical wavelengths (e.g. Fabian \&
Iwasawa 1999). The contribution of Type II QSOs  could be
significant. For instance, Barger \etal (2001) calculate lower and
upper limits of $6 \times 10^4$ and $9 \times 10^5$ $\msun$ Mpc$^{-3}$
for the mass density in the SBHs which comprise the X-ray
background. Gilli, Salvati \& Hasinger (2001) find that the  spectral
shape of the hard (2$-$10 Kev) X-ray background can be best explained
if obscured AGNs evolve more rapidly as a function of redshift than do
their unobscured counterparts. Their model assumes a ratio between
absorbed and unabsorbed AGNs increasing from $\sim 4$ in the local
universe to $\sim 10$ at $z\sim 1.3$, and remaining constant at higher
redshifts. Such a model, if correct, would translate into an increase
by nearly a factor of 10 in the SBH cumulative mass density derived
above.

\section{Black Hole Demographics: Local AGNs}

In the supermassive black hole business, the masses which are most
challenging to measure are those of the black holes powering local
AGNs. Compared to QSOs, lower-luminosity AGNs have a small ratio
between nuclear non-thermal and stellar luminosity, making it
difficult to assess what fraction of the total luminosity is due to
accretion onto the central black hole. Furthermore, the history of
past activity is not known, so it is not obvious what fraction of the
SBH mass, $\mh$, predated the onset of the present nuclear activity
(and the assumption that AGNs radiate at the Eddington limit is not
justifiable). In other words, Soltan's arguments, which hold rather
nicely for high redshift powerful QSOs, are not applicable to their
less flamboyant, nearby cousins.

To make matters worse, the techniques that allow us to detect
supermassive black holes in quiescent galaxies are seldom applicable
to the hosts of AGNs.  In Seyfert 1 galaxies, and in the handful of
QSOs for which traditional dynamical studies of the gas or stellar
kinematics can be  performed, the presence of the bright non-thermal
nucleus (e.g. Malkan, Gorjian \& Tam 1998) overwhelms the very
spectral features which are necessary for dynamical studies.  The only
Seyfert galaxy in which a SBH has been detected by spatially-resolved
kinematics is NGC 4258, which is blessed with the presence of an
orderly water maser disk (Watson \& Wallin 1994; Greenhill \etal 1995;
Miyoshi \etal  1995).  The radius of influence of the black hole at
its center, $\sim$ 0\Sec15, can barely be resolved by the Hubble Space
Telescope (HST) but is fully sampled by the VLBA at 22.2 GHz.
Unfortunately, water masers are rare (Greenhill \etal 2002) and of the
handful that are known, only in NGC 4258 are the maser clouds
distributed in a simple geometrical configuration that exhibits clear
Keplerian motion around the central source (Braatz \etal 1996;
Greenhill \etal 1997, 1996; Greenhill, Moran \& Hernquist 1997;
Trotter \etal 1998).  A study of  black hole demographics in AGNs must
therefore proceed through alternative techniques.

To my knowledge, the only attempt at deriving a mass function  for
local ($z < 0.1$) AGNs was published by Padovani \etal (1990) using
the CfA magnitude limited sample of Seyfert 1 galaxies. For each
galaxy, the mass of the central SBH was derived  from the dynamics of
the broad line region (BLR) under the  virial approximation, $\mh =
v^2 r / G$. The radius $r$ of the BLR was calculated by assuming that
the ionization parameter $U$ (or more precisely, its product with the
electron density) is known and invariant  from object to object. $U$
depends on the inverse square of $r$, and linearly on the number of
ionizing photons; the latter quantity can be derived if the spectral
energy distribution of each object is known. As noted by Padovani
\etal,  however, the ionization parameter is likely {\it not}
invariant. More recently, Wandel, Peterson \& Malkan (1999) have
calibrated the ``photoionization method'' against the more
sophisticated technique which has become known as ``reverberation
mapping'' (Blandford \& McKee 1892; Peterson 1993; Netzer \& Peterson
1997; Koraktar \& Gaskell 1991). The latter method  relies on the fact
that if the non-thermal nuclear continuum is variable, then the
responsivity-weighted radius $r$ of the BLR is measured by the
light-travel time delay between emission and continuum variations.  As
in the case of the photoionization method, the mass of the central
black hole follows from the virial approximation, {\it if} the BLR is
gravitationally bound. The latter assumption has now received strong
support in a few well-studied cases (Koratkar \& Gaskell 1991; Wandel,
Peterson \& Malkan 1999; Peterson \& Wandel 2000, but see also Krolik
2001).

Wandel, Peterson \& Malkan (1999) concluded that photoionization
techniques and reverberation-mapping estimates of the BLR sizes (or
central masses) compare well, but only in a {\it statistical}
sense. In other words, $\mh$ estimates from the two methods can differ
by up to an order of magnitude for individual objects, yet there does
exist a reasonably good linear correlation between the two quantities
when large samples are investigated, which bodes well for the Padovani
\etal analysis. Eight of the Seyfert 1 galaxies in the Padovani \etal
sample also have reverberation-mapping masses (Wandel \etal 1999;
Kaspi \etal 2000). For these galaxies, the reverberation masses are a
factor 3.6$\pm$3.4 larger (in the mean) than the photoionization
masses. In Fig. 1, the cumulative mass function in local Seyfert 1
galaxies derived by Padovani \etal, once corrected for this factor and
scaled to $H_0 = 75$ \kms~ Mpc$^{-1}$, is compared to the mass
function in QSO black holes.

The total density of SBHs in Seyfert 1 galaxies is $\sim 5000$ $\msun$
Mpc$^{-3}$. Despite the upward revision  by a factor $\sim$ eight
compared to the original estimate, the main conclusion reached by
Padovani \etal still holds: ``the bulk of the mass related to the
accretion processes connected with past QSO activity does not reside
in Seyfert 1 nuclei. Instead, the remnants of past activity must be
present in a much larger number of galaxies''.  In the local universe,
the ratio of Seyfert 2 to Seyfert 1 galaxies is $\sim$ four (Maiolino
\& Rieke 1995), while  LINERs are a factor of a few more numerous than
Seyferts (Vila-Vilaro 2000). Yet even after correcting the mass
density given above to include these classes of AGNs, the total
cumulative mass density in local AGNs falls a factor of several below
that estimated for high redshift QSOs. Thus, in the search for SBHs,
powerful QSOs and completely quiescent galaxies appear to be equally
promising targets.

\section{Black Hole Demographics: Local Quiescent Galaxies}

\subsection{Supermassive Black Holes and their Host Galaxies}

Measuring SBH masses in the nuclei of ``normal'' nearby galaxies has
been a staple of the astronomical literature since the late 1970s. It
all started in 1978, when Wallace Sargent  and collaborators published
an investigation of the nuclear dynamics of the Virgo cluster cD, M87
(Sargent \etal 1978), claiming the detection of a five billion solar
mass black hole. Other famous detections followed: M32 in the mid '80s
(Tonry 1984), M31 a few years later (Kormendy 1988).  Each claim,
however, seemed to have its detractors, beginning with  Binney \&
Mamon (1982) who dismantled Sargent's M87 black hole detection and
alerted the community to the perils of the now familiar
``mass-to-light ratio -- velocity anisotropy degeneracy''.

\begin{figure}[t]
\psfig{figure=fig2.epsi,height=2.1in}
\caption{(left) Correlation between central velocity dispersion and
black hole mass for all secure SBH detections. Published data  are
shown as solid symbols, data based on unpublished analyses as open
symbols.\newline Figure 3: (right) Correlation between bulge $B-$band
magnitude and black hole mass for the same sample shown in
Fig. 2. Elliptical  galaxies are shown as circles, lenticulars and
compact ellipticals  as squares, and spirals as triangles}
\end{figure}

Important ground-based work on SBHs continued through the '90s
(Richstone, Bower \& Dressler 1990; Kormendy \etal 1996a, 1996b;
Magorrian \etal 1998), producing a series of tantalizing but
frustratingly inconclusive results. Indeed, it was not until the
launch of HST that dramatic progress was made.  It was HST data that
firmly established the existence of a SBH in M87 (Harms \etal 1994),
thereby ending a two-decade controversy. Since then, SBH masses based
on HST/FOS and STIS data have been published for ten additional
galaxies (Ferrarese, Ford \& Jaffe 1996; Bower \etal 1998; van der
Marel \& van den Bosch 1998; Ferrarese \& Ford 1999; Emsellem \etal
1999; Cretton \& van den Bosh 1999; Verdoes Kleijn \etal 2000;
Gebhardt \etal 2000; Joseph \etal 2001; Barth \etal 2001; Sarzi \etal
2001).

The success of HST can be ascribed to the fact that  its unprecedented
spatial resolution (in the optical regime, at least!) makes it
possible to resolve, in favorable cases, the region of space within
which the SBH's gravitational potential dominates that of  the
surrounding stars, i.e., the ``SBH sphere of influence''.  This is
more crucial than might at first be realized. It has become obvious
(Ferrarese \& Merritt 2000)  that resolving the sphere of influence
does not simply aid the SBH detection: it is a {\it necessary}
condition for a detection to be made.  Ground-based observations
generally lack the spatial resolution necessary to penetrate the SBH
sphere of influence, and this condition inevitably leads to spurious
detections and overestimated masses (see Merritt \& Ferrarese 2001c
for a more thorough discussion of this issue). To date, with a few
notable exceptions (the Milky Way, Genzel \etal 2000, Ghez \etal 2000;
NGC 4258, Miyoshi \etal 1995; NGC 5128, Marconi \etal 2001), all firm
SBH detections ---  {\it detections based on data which resolve the
SBH sphere of influence} --- are based on HST data (see Table 1 of
Merritt \& Ferrarese 2001c).

It was by isolating these secure detections that it became possible to
unveil the existence of a fundamental, seemingly perfect correlation
between black hole mass, $\mh$, and velocity dispersion, $\sigma$, of
the host bulge (Fig. 2, Ferrarese \& Merritt 2000; Gebhardt \etal
2000): the  relation emerged from what had appeared almost as a
scatter plot when the sample was restricted only to galaxies in which
the SBH sphere of influence had been resolved (Ferrarese \& Merritt
2000; Merritt \& Ferrarese 2001b; Merritt \& Ferrarese 2001c). A
regression analysis, accounting for errors in both coordinates, of all
published SBH detections (listed in Table 1 of Merritt \& Ferrarese
2001c) gives

\begin{equation}\mh = (1.66 \pm 0.32) \times 10^8 ~\msun~  {\left({\sigma} \over {200 {\rm ~km~s^{-1}}}\right)}^{4.58 \pm
0.52}.\end{equation}

Including the few preliminary masses based on unpublished analyses produces an
indistinguishable slope of 4.64$\pm$0.47. The reduced $\chi^2$ of the
fit, 0.74, points to a relation with negligible intrinsic scatter, in
agreement with the  observations made by Ferrarese \& Merritt (2000)
based on the smaller sample available at the time. Because of its
tightness, the $\ms$ relation has largely supplanted the well known
correlation between $\mh$ and bulge magnitude $M_B$ (Fig. 3; Kormendy
\& Richstone 1995; Magorrian 1998), and has emerged as the tool of
choice in the study of SBH demographics (Merritt \& Ferrarese
2001b). Indeed, the tightness of the $\ms$ relation is its most
puzzling feature,  presenting formidable challenges to  theoretical
models for the formation and evolution of SBHs (e.g Haehnelt \&
Kauffmann 2000; Kauffmann \& Haehnelt 2000; Adams \etal 2001; Ciotti
\& van Albada 2001; Burkert \& Silk 2001). Even if one assumes that a
tight relation was imprinted at an early stage of galaxy/SBH
formation, it is difficult to understand how it could have survived
unaltered in the face of mergers.  It is especially remarkable that
the relation should hold true for galaxies of disparate Hubble types
(from SBs to compact ellipticals to cDs) belonging to wildly different
environments (from rich clusters to the field), showing perfectly
smooth (e.g. NGC 6251) or highly disturbed (e.g. NGC 5128)
morphologies.  For instance, it has recently been noted (McLure \&
Dunlop 2001) that  the large  scatter in the $\ml$ relation (a reduced
$\chi^2$ of 23, Ferrarese \& Merritt 2000) can be significantly
reduced, but only at the expense of  excluding most lenticular and
spiral galaxies (and the odd elliptical, cf. Fig. 3). Even the
remarkably tight correlation discovered by  Graham \etal (2001)
between $\mh$ and ``concentration parameter'' $C$ (the fraction of
total bulge light contained within a predetermined radius) is marred
by the occasional outlier.  However, {\it every} galaxy, even the ones
which do not obey the $\ml$ or $\mh - C$ relations, seems to conform
magically to the $\ms$ relation.

\subsection{Two Routes to SBH Demographics in Local Quiescent Galaxies}

Because of its tightness, the $\ms$ relation provides us with a direct
and powerful tool to  estimate the mass density of SBHs,
$\rho_{\bullet}$, in the local universe. One approach is to combine
the known mass density of spheroids (e.g. Fukugita \etal 1998) with
the mean ratio between the mass  of the SBH and that of the host
bulge. Merritt \& Ferrarese (2001a) used the $\ms$ relation to
estimate $\mh$ for a sample of 32 galaxies for which  a dynamical
measurement of the mass of the hot stellar component was available
(from Magorrian \etal 1998). For this sample, the frequency function
$N[\log(\mh/M_{\rm bulge})]$ is well approximated by a Gaussian with
$\langle\log(\mh/M_{\rm bulge})\rangle \sim -2.90$ and standard
deviation $\sim 0.45$. This implies $\mh/M_{\rm bulge} \sim 1.3 \times
10^{-3}$ or, when combined with the mass density in local spheroids
from Fukugita \etal (1998), $\rho_{\bullet} \sim 5 \times 10^5$. This
estimate is a factor of five smaller than obtained by Magorrian \etal
(1998) using what we now believe to be inflated values for the masses
of the central black holes in many galaxies.

Here I will present an independent derivation of $\rho_{\bullet}$
which, while not directly leading to $\langle\mh/M_{\rm
bulge}\rangle$, has the advantage of producing an analytical
representation of the cumulative SBH mass density as a function of
$\mh$. The idea is simple: if $\mh$ correlates with the luminosity of
the host bulge, the SBH mass density can be calculated once the
luminosity function of bulges is known. Black hole masses are related
to bulge luminosity directly through the $\ml$ relation, a
representation of which is given by Ferrarese \& Merritt (2000) as
$\log\mh = -0.36M_B + 1.2$. Unfortunately, the large scatter of the
$\ml$ relation (Fig.  3), combined with the small number of galaxies
on which it is based, makes it impossible to establish whether
elliptical and spiral galaxies follow a similar relation. Indeed, the
observations of McLure \& Dunlop (2001) cast doubts on whether spirals
and lenticulars follow an $\ml$ relation at all.  This is unfortunate
since the galaxy luminosity function  does show a dependence on
morphology (e.g., Marzke \etal 1998), and it is therefore desirable to
conduct the analysis independently for different Hubble types.  An
alternative approach is to derive a relation between $\mh$ and bulge
luminosity by combining the $\ms$ relation (which, given the present
sample, seems independent of the morphology of the host galaxy) with
the Faber-Jackson relation for ellipticals and its equivalent for
spiral bulges. The drawback here is that the Faber-Jackson relation
has large scatter and is ill defined, especially for bulges.

The luminosity function for spheroids can be derived from the
luminosity function of galaxies, generally represented as a Schechter
function, once a ratio between total and bulge luminosity (which
depends on the Hubble type of the galaxy considered) is assumed. The
latter  is adopted from Table 1 of Fukugita \etal (1998). Here, I will
use the galaxy luminosity function derived by Marzke \etal (1998) from
the Second Southern Sky Redshift Survey (SSRS2), corrected to $H_0 =
75$ \kms~ Mpc$^{-1}$ and an Einstein-de Sitter universe. Marzke \etal
derived  luminosity functions separately for E/S0s and spirals, in a
photometric band $B_{SSRS2}$. This band is similar to the Johnson's
$B$-band, where representations of both the $\ml$ relation and the
Faber-Jackson relation exist: $B_{SSRS2} = B + 0.26$ (Alonso \etal
1993). A Schechter luminosity function,

\begin{equation} \Phi(L) {\rm d}L = \Phi_0 {\left({L \over {L_*}}\right)^\alpha e^{-L/L_*}}
{{{\rm d}L} \over L_*},\end{equation}  is then easily transformed into
a SBH mass density if $L = A\mh^k$,

\begin{equation} \Psi(\mh) {\rm d}\mh = \Psi_0 \left({\mh \over {M_*}}\right)^{k(\alpha+1)-1}
e^{-(\mh/M_*)^k} {{{\rm d}\mh} \over M_*},\end{equation} where $\Psi_0
= k\Phi_0$, $M_* = (\beta L_*10^{0.4\times0.26}/A)^{1/k}$, and  $\beta
\equiv L/L_{bulge}$ = 0.23 for spirals and 0.76 for E/S0
galaxies. $\beta$ is the sum of the ratios between bulge to total
$B$-band luminosity for different Hubble types, each weighted by the
fraction of the mean luminosity density contributed by each type (from
Fukugita \etal 1998).

\setcounter{figure}{3}
\begin{figure}[t]
\psfig{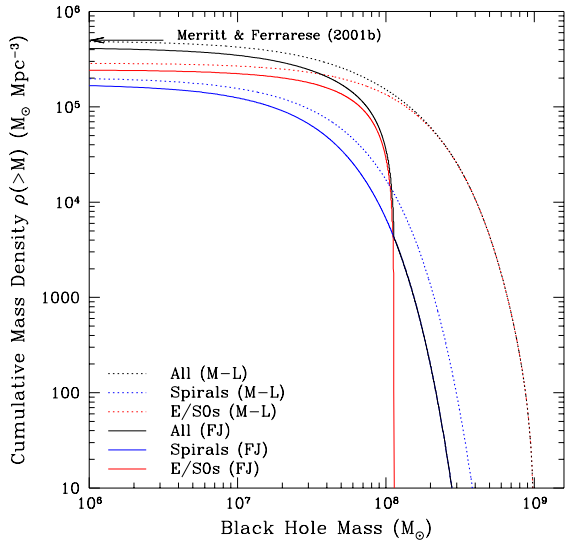}
\caption{The mass function in local black holes for spirals (blue),
E/S0 (red) and a complete  sample of galaxies (black). Dotted lines
are derived from the $\ml$ relation, solid lines  from the $\ms$
combined with the Faber-Jackson relation (as described in the text). }
\end{figure}

Fig. 4 shows the cumulative SBH mass function separately for the E/S0
and spiral populations, derived from the $\ml$ relation (from
Ferrarese \& Merritt 2000, dotted lines) and the $\ms$ relation (from
this paper) combined with the Faber-Jackson relations for ellipticals
and spirals (from Kormendy \& Illingworth 1983, corrected  to $H_0 =
75$ \kms~Mpc$^{-1}$). While the two distributions differ in the
details, there is little difference in the total mass density,  which
falls in the range $(4-5) \times 10^5$ $\msun$ Mpc$^{3}$.  This is in
excellent agreement with the estimate of Merritt \& Ferrarese (2001a).

Table 1 summarizes the mass density estimates for SBHs discussed  in
the preceeding three sections. While a detailed comparison of the
distribution of masses remains to be carried out (for instance, Fig. 1
suggests a larger fraction of very massive black holes, $M > 10^9$, in
high redshift QSOs than have been found in local galaxies), the
overall picture is one of agreement: local studies seem to have
recovered the  overall mass density inferred from high redshift
QSOs. It appears that supermassive black holes are  a fundamental
component of every large galaxy.

\scriptsize
\begin{table}[t]
\begin{tabular}{l c}
\multicolumn{2}{c}{\bf \normalsize Summary of Mass Densities in
Supermassive Black Holes}\\  \hline\hline \multicolumn{1}{c}{Method} &
\multicolumn{1}{c}{$\rho_{\bullet}$ ($10^5$ $\msun$ Mpc$^{-3}$)}\\
\hline   QSO optical counts, $0.3 < z < 5.0$  & $2-4$\\    AGN X-ray
counts, $z > 0.3$ & $0.6 - 9$ \\  Spectral fit to the X-ray
background, $z$ unknown & $2-30$\\   Local AGNs, $z < 0.1$ & $0.05 -
0.6$\\  Local Quiescent Galaxies, $z < 0.0003$ & $4-5$\\ \hline\hline
\end{tabular}
\end{table}
\normalsize

\section {Moving Forward: Open Issues}

\subsection{What More Can Be Learned about the $\ms$ Relation?}

With so much progress in the past few years, it is only natural to be
optimistic about what the near future might bring. Indeed, a
considerable amount of effort will be devoted to the study of
supermassive black holes in nearby galaxies, with HST remaining the
instrument of choice for the investigation. Roughly $130$ galaxies
have, or will be, observed with HST/STIS within the next year. While
only a fraction of these observations are likely to lead to secure SBH
detections (Merritt \& Ferrarese 2001c), these results are highly
anticipated, and will help to better define the slope and scatter of
the $\ms$ relation.

Nevertheless, one important section of  parameter space will remain
unexplored.  Now that the existence of  SBHs is as well established as
that of the galaxies in which they reside, the most pressing need has
become, in my opinion, an exploration of the low mass end of the $\ms$
relation.  However, the vast majority of the galaxies in the HST
pipeline are expected to host SBHs with $\mh \sim 10^8\msun$, a range
already well-sampled by the current data. None of the ongoing programs
is likely to measure a SBH of $\mh<10^7$ $\msun$ (Merritt \& Ferrarese
2001c).

This is unfortunate since determining how far the $\ms$ relation
extends is key for discriminating between different scenarios for the
formation of SBHs. The smallest nuclear SBHs whose masses have been
established dynamically are in the Milky Way (Genzel \etal 2000) and
M32 (Joseph \etal 2001), both with $\mh\approx 3\times10^6\msun$
(Fig. 2). Evidence for black holes with $10^3 < \mh < 10^{6}$ $\msun$
(dubbed ``intermediate'' mass black holes, or IBHs) is so far
circumstantial, the most likely candidates being the super-luminous
{\it off-nuclear} X-ray sources (ULXs) detected by Chandra in a number
of starburst galaxies (Fabbiano \etal 2001; Matsumoto \etal  2001).

The link between IBHs and SBHs is unclear.  If Chandra's off-nuclear
ULXs are indeed IBHs, they could sink slowly to the galaxy center
through dynamical friction and provide the seeds for nuclear SBHs
(Ebisuzaki \etal 2001). Or the latter might be born {\it in situ},
through collapse of a protogalactic cloud, possibly before the onset
of star formation in the bulge (Loeb 1993; Silk \& Rees 1998;
Haehnelt, Natarajan \& Rees 1997).   Deciding between these and
competing formation scenarios will undoubtedly keep theorists busy for
many years.  However, different theories would almost certainly make
different predictions about the form of the $\ms$ relation, and this
is the most promising route for distinguishing between them.  For
instance, {\it in situ} formation in nuclei is unlikely to  result in
black holes less massive than $\sim 10^6\msun$ (e.g Haehnelt,
Natarajan \& Rees 1998), while accumulation of IBHs would probably not
result in as tight a correlation between $\mh$ and $\sigma$ unless
some additional feedback mechanism were invoked (e.g. Burkert \& Silk
2001).  But little progress is likely to be made until we know whether
IBHs are present in galaxy nuclei and if so,  where they lie relative
to the $\ms$ relation defined by SBHs. Therefore, exploring the $\ms$
relation  in the $M < 10^6$ $\msun$ range will be an important
challenge in the years  to come.

A first step in this direction has been taken recently with the
derivation of an upper limit, of a few thousand solar masses, for the
putative black hole inhabiting the nucleus of the nearby spiral M33
(Merritt, Ferrarese \& Joseph 2001; Gebhardt \etal 2001; Valluri \etal
2002). As small as this upper limit might seem, it is still consistent
with the $\ms$ relation as characterized in this paper, when
extrapolated (by three orders of magnitude!) to the thousand solar
mass range. Unfortunately, until the next technological leap, there is
little hope of significantly tightening this upper limit: at the
distance of M33, the black hole's sphere of influence is well below
(by at least an order of magnitude) the resolution capabilities of
HST. Indeed, with one notable exception, there are {\it no} galaxies
expected to contain a black hole below the $10^6$ $\msun$ mark that
are close enough, and have high enough central surface brightness,  to
allow HST to measure $\mh$. The one exception, the Local Group
spheroidal galaxy NGC 205, is scheduled to be observed by HST as part
of program 9448 (P.I. L. Ferrarese). NGC 205 is expected to host a
$\sim 7.5\times 10^5$ $\msun$ black hole; at a distance of 740 kpc, a
black hole as small as $6 \times 10^5$ $\msun$ can be detected.  Even
so, it seems inevitable that, to fully characterize the low mass range
of the $\ms$ relation, we must look beyond HST.

In my opinion, the answer is reverberation mapping. Although the
obvious drawback is that it is only applicable to the 1\% of galaxies
with Type 1 AGNs, reverberation mapping is intrinsically unbiased with
respect to black hole mass, provided the galaxies can be monitored
with the appropriate time resolution. Furthermore, reverberation
mapping can probe galaxies at high redshifts and with a wide range of
nuclear activity, opening an avenue for the exploration of possible
dependences of the $\ms$ relation on cosmic time and activity level.

The stage is being set to embark upon this new endeavor. In the past
few years, the reliability of reverberation-mapping masses has been
called into question on both observational (e.g. Ho  1999; Richstone
\etal  1998) and theoretical (Krolik 2001) grounds. However, on the
observational side, the doubts appear to be dissipating. The
observation that SBHs in AGNs appeared to be undermassive, by a factor
$\sim 50$, compared to SBHs in quiescent galaxies (Wandel 1999), was
apparently the result of two erroneous assumptions: the overestimate
(by a factor $\sim$ six) of SBH masses in quiescent galaxies derived
from the $\ml$ relation of Magorrian \etal (1998); and an overestimate
of the AGN host bulge magnitudes (by up to $\sim 3.5$ mag) adopted by
Wandel (McLure \& Dunlop 2000; Merritt \& Ferrarese 2001c; Wandel
2002). Indeed, Merritt \& Ferrarese (2001c) conclude that  the ratio
of SBH to bulge mass in Seyfert, QSO and quiescent galaxy samples are
all consistent: $\langle \mh/\mb\rangle=0.09\%$ (QSOs) and $0.12\%$
(Seyferts), $\langle\mh/\mb\rangle=0.13\%$ for quiescent galaxies.

On the theoretical side, Krolik (2001) argues that the unknown BLR
geometry, radial emissivity distribution, and angular radiation
pattern of the line emission, coupled with the often less than optimal
temporal sampling of the data, can lead to systematic errors in the
reverberation masses of a factor $\sim$ three or more. While there is
little doubt that Krolik's objections are all well-justified, my
collaborators and I have taken an observational approach to this
issue. Since there are no independent measurements of $\mh$ for any of
the reverberation-mapped AGNs, we have opted for an indirect
comparison by placing these galaxies onto the $\ms$ plane. Initial
results (Ferrarese \etal 2001) suggest that the AGN sample follows the
same $\ms$ relation as the quiescent galaxies on which the relation is
defined. More secure conclusions should be reached within the next
year, once the AGN sample is doubled (Pogge \etal 2002). At the
moment, the evidence suggests that reverberation mapping works, in
spite of the theoretically motivated concerns.

\subsection{Beyond the $\ms$ Relation: Exploring the Dark 
Side of Galaxies}

The $\ms$ relation probes a direct connection between SBHs and
galactic bulges. The velocity dispersion, $\sigma$, is measured within
a region which, though large compared to the black hole sphere of
influence, is at least an order of magnitude smaller than the optical
radius of the galaxy, and is likely dominated by luminous matter
(Faber \& Gallagher 1979). Therefore, $\sigma$ is unable to tell us
about the connection between SBHs and other fundamental baryonic
structures, such as the galactic disk or halo, while the link to the
dark matter (DM) component also remains utterly unexplored.

That this issue has not yet been addressed is somewhat surprising,
since it is not the mass of the bulge but rather, the {\it total mass}
of the galaxy (or of the DM halo), which is the key ingredient of most
theoretical models proposed for the formation of SBHs (Adams, Graff \&
Richstone 2000; Monaco \etal 2000; Haehnelt, Natarajan \& Rees 1998;
Silk \& Rees 1998; Haehnelt \& Kauffmann 2000; Cattaneo, Haehnelt \&
Rees 1999; Loeb \& Rasio 1994). Once the models predict a correlation
with total mass (or DM halo mass), the correlation with bulge mass is
implicit because, in standard CDM scenarios, the bulge mass is loosely
determined by the halo properties (e.g. van den Bosch 2000; Haehnelt,
Natarajan \& Rees 1998; Zhang \& Wyse 2000).

\begin{figure}[t]
\psfig{figure=fig4.epsi,height=2.3in}
\caption{(left) Correlation between the  rotational velocity and bulge
velocity dispersion  for a sample of 16 spiral galaxies (solid
circles) and 21 ellipticals (open circles; plot adapted from Ferrarese
2002).\newline Figure 6: (right) Same as Fig. 5, but with $v_c$ and
$\sigma$ converted to halo mass and black hole mass respectively
(see text for further details). The upper limit on the SBH mass in M33
(Merritt \etal 2001) is shown by the arrow.}
\end{figure}

It is  natural to ask whether the $\ms$ relation might just be the
by-product of an even more fundamental relation between  $\mh$ and the
total gravitational mass of the galaxy. As it turns out, such a
fundamental relation is  likely to exist (Ferrarese 2002). Fig. 5
demonstrates the existence of a tight correlation between the bulge
velocity dispersion (the same quantity used in defining the $\ms$
relation, typically measured within an aperture of size $R \lae 0.5$
kpc) and the circular velocity $v_c$, measured at radii $R \sim 20-80$
kpc, for a sample of 16 spiral galaxies.
A regression analysis, accounting for errors in both variables, gives

\begin{equation}\log{v_c} = (0.88 \pm 0.17) \log{\sigma} + (0.47 \pm 0.35)\end{equation}
with a reduced $\chi^2$ of 0.64.

For spiral galaxies, $v_c$ is measured directly from HI or optical
rotation curves. In elliptical galaxies, $v_c$ can be derived from
dynamical models of the observed stellar absorption line profiles,
velocity dispersion and surface brightness profiles. Fig. 5 shows that
the spirals naturally blend with a sample of 21 elliptical galaxies
(from Kronawitter et al. 2000) in the $\vs$ plane; both samples obey
the relation given in equation (5).

The implications of equation (5) are exciting.  The circular velocity
$v_c$ is a measure of gravitational mass through the virial theorem,
and can be related to the DM halo mass (Navarro \& Steinmetz 2000;
Bullock et al. 2001).  Keeping in mind that, as discussed in section
5.1,   the $\ms$ relation is not well defined below $10^7$ $\msun$,
and not defined at all below $10^6$ $\msun$, the $\vs$ relation can be
translated into a relation between the mass of the central black hole
(related to $\sigma$ through equation 2) and that of the DM halo
(Fig. 6):

\begin{equation}{{\mh} \over {10^8~\msun}} \sim 0.046 {\left({M_{DM}} \over {10^{12}~ \msun}\right)}^{1.6}\end{equation}

\noindent (Ferrarese 2002). The existence of this relation seems to
conflict with recent claims that SBHs do not relate to any other
galactic structure but the bulge (Richstone 1998; Kormendy \& Gebhardt
2001; Gebhardt \etal 2001).

The relation between $\mh$ and $M_{DM}$ is non-linear, with the ratio
$\mh/M_{DM}$ decreasing from $6\times10^{-5}$ for $M_{DM} \sim 
10^{14}$ $\msun$, to $5\times10^{-6}$ for  $M_{DM} \sim 10^{12}$
$\msun$. Haehnelt, Natarajan \& Rees (1998) advocated a nonlinear
relation between SBH and DM halo mass in order to reproduce the
luminosity function of QSOs, noting that a linear relation would
translate into too low a value for the QSO duty cycle, $t_{QSO} \sim 3
\times 10^5$ yr. Increasing the QSOs lifetime to values more in line
with current observational constraints (e.g. Martini \& Weinberg 2001)
produces an increasingly steeper relation between $\mh$ and
$M_{DM}$. If $t_{QSO }\sim 1.5 \times 10^7$ yr (equal to the Salpeter
time), then the slope of the $\mh -M_{DM}$ relation must be increased
to $\sim 2$ to provide a reasonable fit to the QSO luminosity
function.  The empirical correlation shown in Fig. 6 seems to support
such claims. Furthermore, Fig. 6 indicates that the tendency of massive halos to become
less efficient in forming SBH as $M_{DM}$ decreases, is even more
pronounced for halos with $M_{DM} < 10^{12}$ $\msun$, and breaks
down completely in the case of M33.  Such halos
might indeed be {\it unable} to form SBH, as proposed  on theoretical
grounds by Haehnelt, Natarajan \& Rees (1998) and Silk \& Rees (1998).

\subsection {Additional Clues to SBH and Galaxy Formation and Further Challenges}

I conclude this review with some general comments about the early
stages of galaxy and black hole formation.  The $\ms$  relation (and,
even more so, the $\mh - M_{DM}$ relation) implies a causal connection
between the evolution of black holes and their host galaxies.  But
what came first: the stars or the black holes? And was the $\ms$
relation  imprinted during the early stages of galaxy formation?  The
answer to the latter question is generally assumed to be affirmative,
but in fact we have no direct proof of it.  The most distant galaxy in
the $\ms$ plot (NGC 6251, Ferrarese \& Ford 1999) is at $\sim 100$
Mpc. Studies of reverberation-mapped galaxies (Ferrarese \etal 2001,
Pogge \etal 2002) have reached two times farther, and it might be
technically possible to push the envelope up to $z \sim 1$.  It seems
unlikely that we will ever be able to build an $\ms$ relation at the
redshift corresponding to the optically bright phase of the QSOs ($z
\sim 2- 3$), let alone at redshifts at which the first protogalactic
fragments are believed to have formed, $z > 5$. Present day dwarf
galaxies might very well be relics from such an era (Mateo 1998;
Carraro \etal 2001); however, detecting SBHs in these systems requires
a spatial resolution well beyond the capabilities of present
instrumentation. In fact, the $\ms$ relation is defined
primarily by bright giant ellipticals which are likely to have an
extensive history of merging.  In other words, we have no direct
information about the ``primordial'' connection between supermassive
black holes and their hosts: what we see is the result of gigayears of
evolution.

\setcounter{figure}{6}
\begin{figure}[t]
\psfig{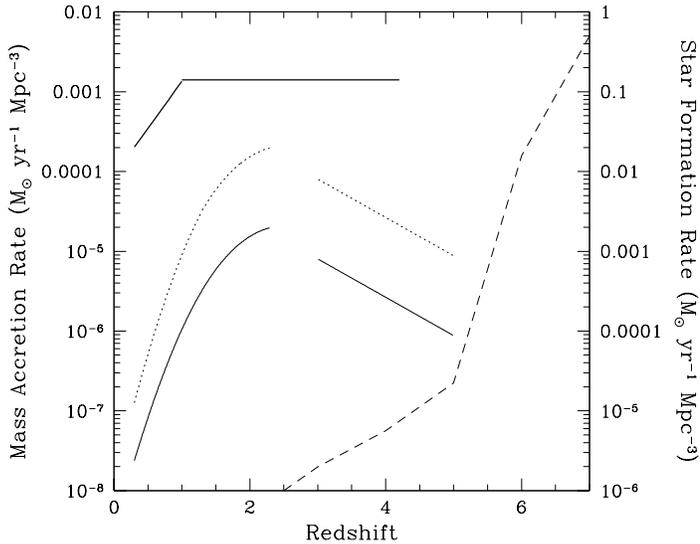}
\caption{The mass accretion rate onto supermassive black holes with
$\mh > 7 \times 10^8$ $\msun$ (thin solid line, units shown on the
left axis), compared  to the star formation rate from Steidel \etal
1999 (thick solid line,  units on the right axis) and the epoch of
formation of the DM halos  which host such SBHs according to Fig. 5
and 7 (dashed line, from  Gottl\"{o}ber, Klypin \& Kravtsov
2001). Eighty percent of the halos which (presumably) host QSO engines
are capable of forming  stars at redshifts $z > 5$. The dotted line
shows the mass accretion rate  onto high redshift QSOs corrected for
the contribution of obscured objects,  as in Gilli \etal 2001.}
\end{figure}

A scenario in which galaxy formation precedes the formation of
supermassive black holes seems to fit more naturally within the
current paradigm of hierarchical structure formation
(e.g. Miralda-Escude \& Rees 1997). For instance, star formation can
proceed in halos with virial temperature as low as $10^4$ K, which can
form at redshifts $z > 10$ (e.g. Ostriker \& Gnedin 1996). Subsequent
stellar evolution in these systems would produce enough energy through
stellar winds or supernovae explosions to expel most of the remaining
gas from the shallow potential wells (Couchman \& Rees 1986; Dekel \&
Silk 1986), likely inhibiting the formation of supermassive black
holes. Deeper potential wells, which are more conducive to SBH
formation (e.g. Haehnelt, Natarajan \& Rees 1998) would only form at
later times.  Studies of elemental abundances in high redshift ($z >
3$) QSOs support this view:  most of the metal  enrichment and star
formation seem to have taken  place at least 1 Gyr before the luminous
phase of the QSO (Hamann \& Ferlan 1999 and references therein;
Dietrich, \etal 2001).

Fig. 7 shows a comparison between the mass accretion rate onto
optically luminous QSOs with $\mh > 7\times10^8$ (corresponding to the
magnitude limit of the SDSS QSO Survey for objects radiating at the
Eddington limit), and the star formation rate from Steidel \etal 1999
(see also Abraham \etal 1999; Cowie \etal 1997).  Similarities between
the two curves, which have been noted many times (e.g. Boyle \&
Terlevich 1998) are diminished by these recent results, even after the
QSO results are corrected for the possible contamination of obscured
objects (Barger \etal 2001; Gilli \etal 2001; Salucci \etal 1998). If
anything, Fig. 7 supports the conclusion that star formation was well
underway by the time the QSOs started shining.

The connection between QSO activity and merging rate is also not
readily apparent: observations show that the merging rate depends on
redshift as $(1+z)^{\alpha}$ with $\alpha = 2 - 4$ (Le Fevre \etal
1999; Burkey \etal 1994; Carlberg \etal 1994; Yee \& Ellingson 1995;
Abraham 1999). Even in the $z < 2.3$ range, where both curves decline,
the number of mergers declines by at most a factor 30, while the
comoving density of QSO declines by three orders of magnitude.
Perhaps more telling is the comparison with the merging history of DM
halos and the ensuing formation of galaxies. Fig. 7 also shows the
distribution of formation redshifts for present day halos with virial
velocities $> 300$ \kms~ taken from the N-body simulation  of
Gottl\"{o}ber, Klypin \& Kravtsov (2001). According to Figures 5 and
6, these are the halos associated with the black holes sampled by the
SDSS, also shown in Fig. 7. Virtually all  such halos are able to host
a luminous galaxy  (a condition reached when the halo progenitor first
reaches a virial velocity $>  50$ \kms) before a redshift $\sim 2.5$,
i.e. before the optically bright phase of the QSOs.

In the midst of all this, one thing is certain: SHBs can no longer be
studied in isolation. Understanding how they form, and how they shape
their surroundings, requires a good deal more information from
seemingly unrelated fields than could have been anticipated just a few
years ago.


\end{document}